\documentclass[10pt]{iopart}

\usepackage{iopams}
\usepackage{graphicx}
\usepackage{url}
\usepackage{siunitx} 
\usepackage{hyperref}
\hypersetup{
    colorlinks=true,
    linkcolor=blue,
    filecolor=magenta,      
    urlcolor=blue,
}

\begin{document}

\title[3D non-linear MHD simulations of core density collapse event in LHD plasma]{3D non-linear MHD simulations of core density collapse event in LHD plasma}

\author{A. Civit-Bertran$^1$, S. Futatani $^{1}$\footnote{Author to whom any correspondence should be addressed}, Y. Suzuki $^2$ and J. Dominguez-Palacios$^3$ }

\address{$^1$ Universitat Politècnica de Catalunya, Avinguda Diagonal 647, 08028 Barcelona, Spain}
\address{$^2$ Graduate School of Advanced Science and Engineering, Hiroshima University, 739-8527
Higashi-Hiroshima, Japan}
\address{$^3$ Fiat Lux, San Diego, CA 92101, United States of America}
\ead{\href{mailto:albert.civit.bertran@upc.edu}{albert.civit.bertran@upc.edu},\href{mailto:shimpei.futatani@upc.edu}{shimpei.futatani@upc.edu}}
\vspace{10pt}
\begin{indented}
\item[]July 2025
\end{indented}

\begin{abstract}A new three-dimensional, non-linear Magnetohydrodynamics (MHD) model has been extended in MIPS code, incorporating parallel heat diffusivity. The model has been benchmarked against the former MHD model used in MIPS code. A preliminary study of the core density collapse event (CDC) observed in the Large Helical Device (LHD) plasma has been performed using the developed model.  The equilibrium has been constructed using HINT code for a typical super dense core discharge in LHD, with vacuum magnetic axis configuration $R_{\rm axV} = \SI{3.85}{\, m}$ and magnetic axis beta  $\beta_{0}=4\%$ plasma. This configuration corresponds to a plasma with a steep pressure gradient and strong Shafranov shift, which makes the plasma potentially unstable in the LHD. The model shows preliminary characteristics of the CDC event. The plasma is destabilized by high-$n$ ballooning modes in the low-field side region during the linear regime, eventually leading to the collapse of the pressure and density profiles, together with the stochastization of the magnetic field and a shift to low-$n$ modes centered at the core of the plasma after the non-linear coupling at the relaxation regime. 
\end{abstract}

%
\vspace{2pc}
\noindent{\it Keywords}: 3D non-linear MHD simulations, core density collapse event, stellarator. 
%
\submitto{\PPCF}
%
%
\ioptwocol

\section{Introduction}
In the Large Helical Device (LHD) experiment, super dense core (SDC) plasma discharges can be achieved in ``standard'' and outward-shifted magnetic configurations (vacuum magnetic axis $R_{\rm axV} \geq \SI{3.75}{\, m}$) by means of multiple consecutive pellet injections. The SDC discharges have been achieved in both Local Island Divertor (LID) \cite{Ohyabu_2006, Morisaki_2007} and Helical Divertor (HD) configurations \cite{Yamada_2007, Sakamoto_2007, Sakamoto_2008}. This high performance configuration is characterized by peaked pressure and density profiles with steep gradients, while electron temperature has a broad profile and is considerably low, $T_{\rm e}  < \SI{1}{\, keV}$. These profiles can be sustained by means of an internal diffusion barrier (IDB) which separates the plasma into different regions: SDC region inside the IDB (core of the plasma), the IDB region and the mantle region outside the IDB. The central pressure increase is limited by a MHD instability that causes a flushing of central pressure and density \cite{ Miyazawa_2008, Ohdachi_2008}. 
The steep pressure gradient and strong Shafranov shift induce the MHD instability, which is in the order of the sub-millisecond timescale. The instability is characterized by a drop in the central pressure accompanied by a drop in the central density, whilst the central temperature remains practically unaffected. This phenomenon is referred to as the core density collapse (CDC) event. Oscillations on the outboard side of the plasma were found before the collapse, which were consistent with the prediction of ballooning modes \cite{Ohdachi_2010, Ohdachi_2017, Mizuguchi_2008, Kinoshita_2022}. The aim of the study is to replicate the non-linear analysis of the instabilities observed in a typical IDB-SDC discharge to further understand the dynamics driving the CDC event.

In this work, a new $3$D non-linear MHD model has been developed in MIPS code \cite{Todo_2010} including parallel heat conductivity. The developed MHD equations represent the time evolution of plasma density $(n)$, momentum $(\rho\mathbf{v})$ and single fluid temperature $(T=T_{\rm i}+T_{\rm e}$ $\approx 2T_{\rm e})$. The approach allows to analyze the density and the temperature dynamics, including anisotropic plasma heat conductivity. To analyze the process of the CDC event, a 3D MHD equilibrium with a steep pressure profile and strong Shafranov shift has been prepared using the HINT code \cite{Suzuki_2006, Suzuki_2017}. The equilibrium profile mimics a typical IDB-SDC LHD discharge where the CDC event is observed. The non-linear MHD evolution is computed using the MIPS code based on the equilibrium produced by HINT code. In this manuscript, the characteristics of the CDC event in LHD plasma have been analyzed.

The manuscript is organized as follows. The equilibrium profile and its characteristics are discussed in Section \ref{Sec:3D-equilibrium}. In Section \ref{Sec:model-equations}, the new set of equations is introduced, and the benchmark of the new model is discussed in Section \ref{Sec:Benchmark}. In Section \ref{Sec:simulation-results}, the analysis of the CDC event is discussed. Conclusions and perspectives are summarized in Section \ref{Sec:Conclusions-perspectives}.

\section{3D equilibrium of LHD plasma} \label{Sec:3D-equilibrium}
The 3D full-torus MHD equilibrium of LHD plasma is constructed using HINT code, which considers the stochastic regions of the magnetic field and the existence of magnetic islands. The LHD device has $l=2$ poloidal winding number and $M=10$ toroidal field periods. The vacuum magnetic axis can be set to values ranging $R_{\rm axV} = 3.5-\SI{3.9}{\, m}$, with averaged minor radius $a\sim\SI{0.6}{\, m}$, and magnetic field strength $B\leq \SI{3}{\, T}$. Each periodic region spans an angle of $2\pi/10$ radians. The toroidal angle $\phi = 0$ corresponds to the vertically elongated poloidal slice, while $\phi = (1/2)(2\pi/10)$ corresponds to the horizontally elongated poloidal slice. 

The modeled 3D equilibrium mimics a typical LHD shot where CDC event is observed: the outward-shifted magnetic configuration $R_{\rm axV} = \SI{3.85}{\, m}$, with axis toroidal magnetic field $B_t > \SI{2.0}{\,T}$, finite-$\beta$ plasma, using intrinsic helical divertor configuration with IDB-SDC, after the last pellet injection at the time of maximum pressure before the CDC event occurs. This configuration is characterized by an axis high density, high pressure, and relatively low temperature. The typical central beta ($\beta_0$) values where CDC event is observed in this configuration ranges for $\beta_0 \in [4\%,5\%]$ \cite{Ohdachi_2017}.

To generate the finite-$\beta$ equilibrium to be analyzed in this study, the magnetic field at the magnetic axis is defined to be $B_{0} =\SI{2.77}{\, T}$, and the finite beta at the axis is prescribed to $\beta_{0} =\SI{4}{}\%$, which is defined as
\begin{equation}
    \beta_0 = \frac{p_0}{B_0^2/2\mu_0},
\end{equation}where $p_0$ and $B_0$ are the pressure and magnetic field at the magnetic axis, respectively. The pressure profile is defined by the equation:
\begin{equation} \label{HINT:pressure_profile}
    p = p_0(1-\Phi_{\rm N})^4,
\end{equation}with $\Phi_{\rm N}$ the normalized toroidal flux. The pressure and magnetic field profiles are evolved following the procedure described in \cite{Suzuki_2006, Suzuki_2009, Suzuki_2020}. This configuration gives an average $\beta$ value of $\langle\beta\rangle \sim 1\%$, where the value is computed as:
\begin{equation}
    \langle\beta\rangle = \frac{\langle p\rangle}{\langle B^2\rangle/2\mu_0}, \quad \text{with} \quad \langle \cdot\rangle = \frac{\int_{V_{\rm P}} \cdot \, {\rm dV}}{\int_{V_{\rm P}} \, {\rm dV}},
\end{equation} and $V_{\rm P}$ is the plasma volume. 
\begin{figure}[t]
    \centering
    \includegraphics[width=\linewidth]{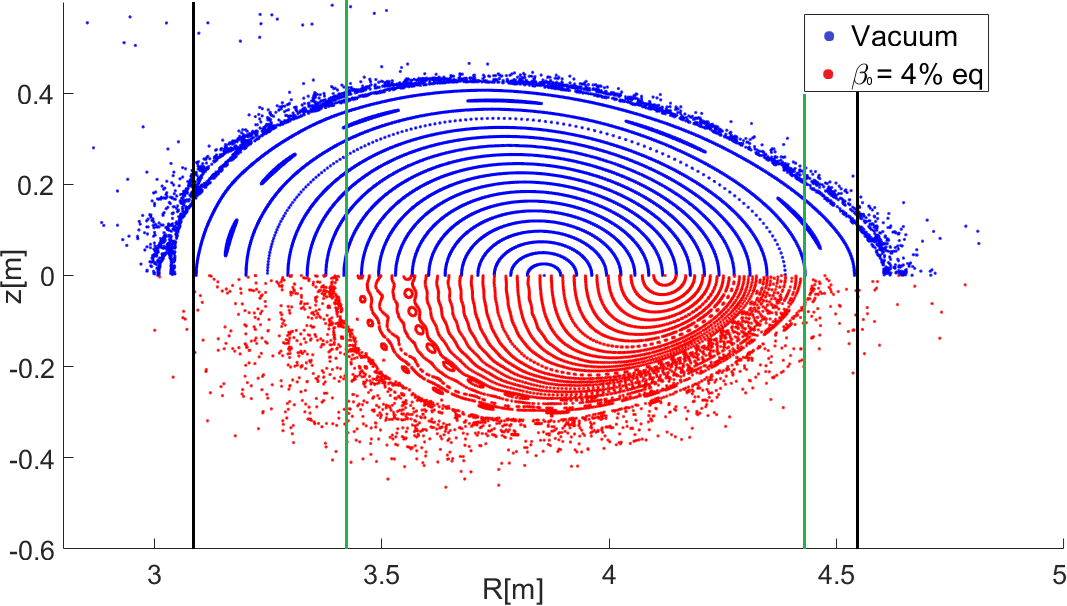}
    \caption{Poincar\'e plot of vacuum (blue) and the IDB-SDC LHD $R_{\rm axV}  = \SI{3.85} {\,m}$ $\beta_0 = 4\%$ equilibrium (red) at the horizontally elongated poloidal slice $\phi = (1/2)(2\pi/10)$. Vertical lines show the radial location of the LCFS at $Z=\SI{0}{\,m}$ of vacuum $R_{\rm axV} = \SI{3.85}{\, m}$ configuration (black), and equilibrium axis $\beta_0 =4\%$ (green).}
    \label{fig:poincare-vacuum-equilibrium}
\end{figure}
Figure \ref{fig:poincare-vacuum-equilibrium} shows the Poincar\'e plot of vacuum and of the IDB-SDC LHD $R_{\rm axV} = \SI{3.85}{\, m}$ $\beta_0 =4\%$ equilibrium at the horizontally elongated poloidal slice $\phi = (1/2)(2\pi/10)$ or $\phi = 18^{\circ}$. After the process of relaxation of the pressure, the equilibrium finite-$\beta$ axis is $R_{\rm ax} = \SI{4.12}{\, m}$. The last closed flux surface (LCFS) of the finite-$\beta$ equilibrium in the high-field side of the plasma, indicated by the left green vertical line, shifts significantly outwards in the major radius direction compared to the LCFS of the vacuum configuration, indicated by the left black vertical line; while in the low-field side of the plasma the shift is inward and less pronounced, indicated by the right green vertical line.

After the convergence of the equilibrium, the pressure profile is cut at $p_{edge} = 0.05 p_{\rm 0}$ for numerical simplicity. This truncation cuts the profile close to the LCFS but slightly outside it; that is, in the region outside of the cut, pressure is defined constant at $p_{edge}$. Thus, the plasma profile does not include the mantle region, as it is beyond the scope of the study, which aims to analyze the core dynamics of the MHD instability. The normalization of the toroidal flux is done at the same position where pressure is cut. A magnetostatic equilibrium, $i.e.$, $\mathbf{v_0} = \mathbf{0}$ is assumed.
The broad temperature profile observed in \cite{Sakamoto_2008,Ohdachi_2017} is approximated by a polynomial function of the normalized toroidal flux as:
\begin{equation} \label{eq:HINT-Temperature}
    \centering
    T \propto T_{0}(1-\Phi_{\rm N}^8)(1-\Phi_{\rm N}^2),
\end{equation}
where $T_{0}$ is the axis temperature and $\Phi_{\rm N}$ the normalized toroidal flux. The first polynomial term ensures an almost flat temperature at the SDC region while keeping a high temperature gradient outside of it; and the second term smooths the boundary regions. The density profile is determined by $p = nT$. 

\begin{figure}[t]
    \centering
    \includegraphics[width=\linewidth]{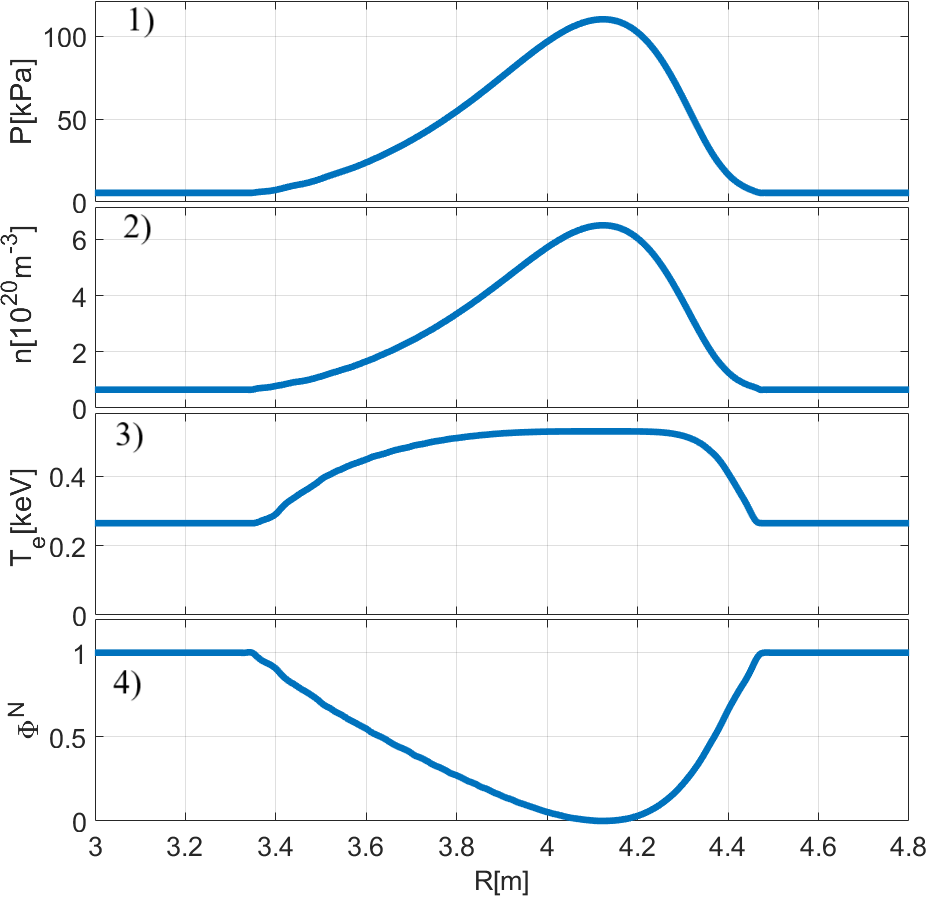}
    \caption{Plasma profiles versus major radius at $Z=\SI{0}{\,m}$ and $\phi = (1/2)(2\pi/10)$ of the IDB-SDC LHD $R_{\rm axV}  = \SI{3.85} {\,m}$ $\beta_0 = 4\%$ equilibrium of 1) pressure, 2) density, 3) electron temperature, and 4) normalized toroidal flux.}
    \label{fig:equilibrium-plasma-profile}
\end{figure}
Figure \ref{fig:equilibrium-plasma-profile} shows the equilibrium plasma profile versus major radius at $Z = \SI{0}{\,m}$ and $\phi = (1/2)(2\pi/10)$ (horizontally elongated poloidal slice) of pressure, density, electron temperature and normalized toroidal flux, as obtained by HINT code calculation. The central pressure, density and electron temperature are: $p_{0} = \SI{110}{\,kPa}$, $n_{0} = \SI{6.5e+20}{\,m^{-3}}$, $T_{\rm e0} = \SI{0.53}{\,keV}$, respectively. At the edge, the values are: $p_{\rm edge} = \SI{5.5}{\,kPa}$, $n_{\rm edge} = \SI{0.65e+20}{\,m^{-3}}$, $T_{\rme,\rm edge} = \SI{0.26}{\,keV}$. The generated equilibrium replicates the broad temperature profile and peaked density and pressure profiles characteristic of the IDB-SDC LHD discharges.   

\begin{figure*}
    \centering
    \includegraphics[width=\linewidth]{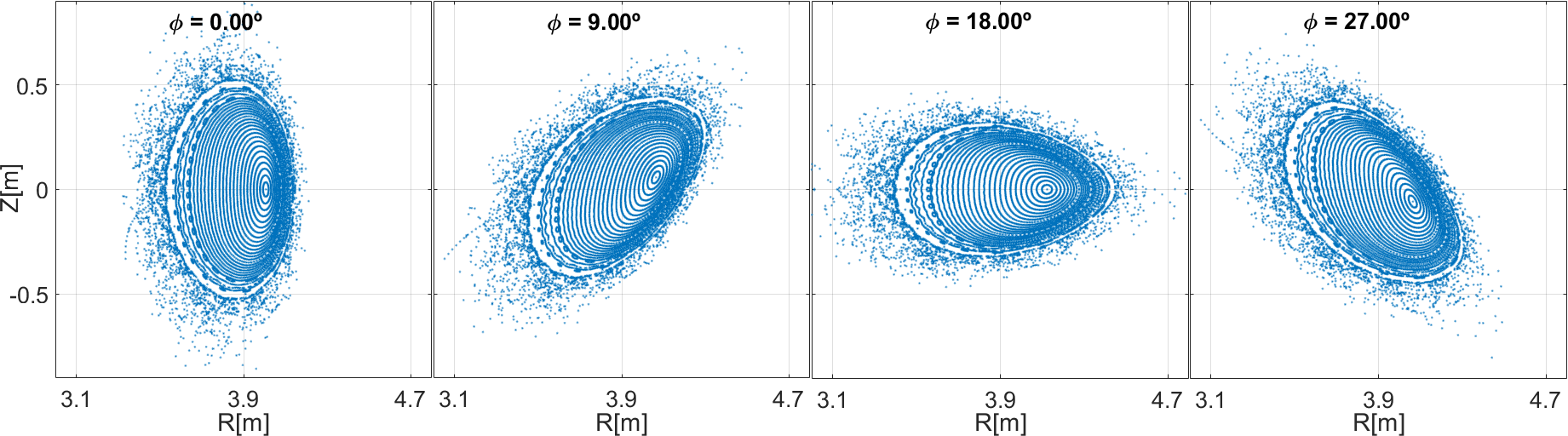}
    \caption{Poincar\'e plot of four different equispaced poloidal slices in the first periodic region $\phi=[0,2\pi/10)$ of the 3D equilibrium of IDB-SDC LHD $R_{\rm axV}  = \SI{3.85} {\,m}$ $\beta_0 = 4\%$ plasma constructed using HINT.}
    \label{fig:equilibrium-poincare}
\end{figure*}
Figure \ref{fig:equilibrium-poincare} shows the Poincar\'e plot of the 3D equilibrium together with the plasma shape variation along the toroidal angle for $4$ different slices, covering an entire periodic region $\phi = [0, 2\pi/10)$. Clear magnetic flux surfaces are observed in the core region. Strong Shafranov shift is observed due to the high beta value. The edge region shows stochastic characteristics.

\section{Model equations and methodology} \label{Sec:model-equations}
A new 3D non-linear MHD model has been extended in MIPS code \cite{Todo_2010}. The developed model solves the evolution of density \textit{$n$}, momentum \textit{$\rho\mathbf{v}$}, single-fluid temperature $T = T_{\rm e} + T_{\rm i} \approx 2T_{\rm e}$, magnetic field $\mathbf{B}$, electric field {$\mathbf{E}$} and current density $\mathbf{J}$:
\begin{eqnarray}
    \label{eq:MHD-n}
    \frac{\partial n }{\partial t} = &- \nabla\cdot(n\mathbf{v}) + \nabla \cdot( D_\perp\nabla n) + S_{\rm n},
\end{eqnarray}
\begin{eqnarray}
    \label{eq:MHD-rho.v}
    \eqalign{\frac{\partial \rho \mathbf{v}}{\partial t} = &-\nabla\cdot(\rho\mathbf{v}\mathbf{v}) - \nabla (n T) + \mathbf{J}\times\mathbf{B} \\
    &+  \nabla \cdot\bigg( \rho\nu \bigg[\nabla \mathbf{v} + (\nabla\mathbf{v})^{\rm T} - \frac{2}{3}(\nabla\cdot\mathbf{v})\mathbb{I}\bigg]\bigg) \\
    &+ \textbf{S}_{\rho\textbf{v}},}
\end{eqnarray}
\begin{eqnarray}
    \label{eq:MHD-T}
    \eqalign{\frac{\partial T}{\partial t}  = &- \nabla\cdot(T\mathbf{v}) -(\gamma-2)T\nabla\cdot\mathbf{v} \\
    &+ \frac{\gamma - 1}{n}\bigg[\nabla \cdot \bigg(\kappa_\perp\nabla_\perp T + \kappa_\parallel\nabla_\parallel T\bigg)\bigg] \\
    &+\frac{\gamma - 1}{n}\bigg[\frac{1}{2}m_iv^2\bigg(\nabla \cdot( D_\perp\nabla n) + S_{\rm n}\bigg)\bigg]\\
    &+\frac{\gamma - 1}{n}\bigg[S_{\rm T} - \textbf{v}\cdot \textbf{S}_{\rho\textbf{v}}\bigg]\\
    &- \frac{T}{n}\bigg[\nabla \cdot( D_\perp\nabla n) + S_{\rm n}\bigg]} 
\end{eqnarray}
\begin{equation}
    \partial_t \mathbf{B} = -\nabla \times \mathbf{E},
\end{equation}
\begin{equation}
    \mathbf{E} = -\mathbf{v}\times\mathbf{B} + \eta\mathbf{J},
\end{equation}
\begin{equation}
    \nabla\times\mathbf{B} = \mu_{\rm 0}\mathbf{J},
\end{equation} 
where $\rho$ is the mass density, $m_i$ is the ion mass, the viscosity is $\nu$, the resistivity $\eta$, particle diffusivity is $D_\perp$ and heat conductivity $\kappa$. Heat conductivity is split into parallel ($\parallel$) and perpendicular ($\perp$) components to account for the anisotropic effects of thermal diffusion. The parallel and perpendicular gradients are defined as $\nabla_\parallel =  \mathbf{b}(\mathbf{b}\cdot \nabla)$ and $\nabla_\perp = \nabla - \nabla_\parallel$, where $\mathbf{b} = \mathbf{B}/|\mathbf{B}|$. The adiabatic constant is $\gamma = 5/3$ and $\mu_0$ is the vacuum permeability. $S_{\rm n}$, $ \mathbf{S}_{\rho\mathbf{v}}$ and $S_{\rm T}$ are the particle, momentum, and heat sources, respectively. In this work, $S_{\rm n}$, $ \mathbf{S}_{\rho\mathbf{v}}$ and $S_{\rm T}$ are set to 0 for the sake of the scope of the work, which is to characterize the CDC event.
Viscous and ohmic heating are not considered, for simplicity. In this study, the generated equilibrium has a large density in the core and relatively low temperature so that core plasma can be considered as collisional and the single-fluid temperature assumption is valid.

The numerical grid is an equispaced, rectangular cylindrical coordinate grid $(R,\phi, Z)$, where the coordinates span: $R \in [2.8, 5.0]\,  \rm m$, $\phi \in [0, 2\pi)$, $Z\in [-1.1,1.1]\,  \rm m$. Time integration is computed using the $4$th-order explicit Runge-Kutta scheme. The spatial derivatives are computed using the $4$th-order central difference method in each dimension. To alleviate spurious oscillations in the evolution, the Kawamura-Kuwahara scheme ($3$rd-order upwind scheme)\cite{KAWAMURA1986145} is used to solve convection terms. A binary mask is used to differentiate numerically the ``plasma'' region from the ``out-of-plasma" region to solve the evolution. In the region inside the mask, the evolution of $n$, $\rho\mathbf{v}$, $T$, $\mathbf{B}$, $\mathbf{E}$ and $\mathbf{J}$ is solved. Outside the mask, only the evolution of $\mathbf{B}$, $\mathbf{E}$ and $\mathbf{J}$ is considered. The binary mask prevents unexpected numerical instabilities that could occur outside the target area. To define the boundary of the binary mask, an analytical formula of an ellipse is used:
\begin{eqnarray}
    r_{mask} &= r_0 + r_1cos(\theta), \quad z_{mask} &= z_0 + z_1sin(\theta),
\end{eqnarray}
where $r_0 = \SI{3.9}{\, m}$, $r_1 = \SI{0.5}{\, m}$, $z_0 = \SI{0.0}{\, m}$ and $z_1 = \SI{0.9}{\, m}$, and $\theta$ is the poloidal angle. The binary mask is then rotated following the toroidal angle so that it adjusts to the plasma shape at each poloidal slice. The size of the mask is larger than that of the plasma, so that the plasma shape is not fixed and can relax outwards. 
The code uses MPI parallelization (see \cite{Futatani_2019} for scaling of parallel computing performance). The simulations for the benchmark in Section \ref{Sec:Benchmark} used $896$ MPI tasks in Marconi/Leonardo-CINECA HPC machine, and took $\SI{24}{\, h}$ of computation, $i.e.$, $\SI{21504}{\, cpuh}$. The simulation for the CDC study in Section \ref{Sec:simulation-results} used $1792$ MPI tasks and took $\SI{108}{\, h}$ of computation, $i.e.$, $\SI{193536}{\, cpuh}$.

\section{Benchmark of developed model} 
\label{Sec:Benchmark}
The developed model has been benchmarked against the former MHD model in MIPS code \cite{Todo_2010, Futatani_2019}. The former model solved the evolution of $\rho$, $\rho\mathbf{v}$ and $P$, with arbitrary initial plasma profiles. The viscous term in the momentum equation was $\nabla\cdot\mathbf{\Pi} \approx (4/3)\nabla(\rho\nu(\nabla\cdot\mathbf{v}) - \nabla\times(\rho\nu(\nabla\times\mathbf{v})$, and perpendicular heat diffusivity was $\chi_\perp\nabla^2(P-P_{\rm eq})$, that is, the diffusion terms were applied to the perturbation of the plasma.  The developed model solves the evolution of $n$, $\rho\mathbf{v}$ and $T$. The model has been extended to consider anisotropic heat conductivity and plasma parameters  $\eta$, $\nu$ and $\kappa$ with spatial profile depending on temperature as $\nu(T) \propto \nu_{0} (T/T_0)^{-3/2}$,  $\eta(T) \propto \eta_{0}(T/T_0)^{-3/2}$ and $\kappa_\parallel(T) \propto \kappa_{\parallel,0} (T/T_0)^{5/2}$, where $\nu_{0},\eta_{0},\kappa_{\parallel,0}$ are the values at the magnetic axis. The viscous term is extended to $\nabla\cdot\mathbf{\Pi}  = \nabla \cdot (\rho\nu [\nabla \mathbf{v} + (\nabla\mathbf{v})^{\rm T} - \frac{2}{3}(\nabla\cdot\mathbf{v})\mathbb{I}])$ and heat and particle diffusion terms are computed as divergences of fluxes. The diffusion terms are applied to the whole profile of the plasma.

While the previous MIPS model was capable of simulating the CDC event, the model has been extended for the following reasons: 1) it is convenient to replace the pressure evolution equation with temperature evolution to establish a foundation for future multi-fluid model development, 2) although the previous MIPS reference model agreed well with the current model, explicitly including parallel heat conductivity allows a more accurate investigation of magnetic field stochasticity effects on MHD dynamics, 3) incorporating particle, momentum, and heat sources enables the study of plasma behavior influenced by external inputs in future work, 4) finally, the binary mask location has been extended outward compared to prior analyses, reducing boundary condition effects and improving the reliability of plasma core region analysis.

The benchmark has been performed using the equilibrium generated by HINT code of IDB-SDC LHD $R_{\rm axV}  = \SI{3.85} {\,m}$, $\beta_0 = 4\%$ plasma configuration (see Section \ref{Sec:3D-equilibrium}). However, density profile is set to be flat, so that in this case initially the temperature profile has the same shape as pressure. Plasma parameters are uniform throughout the domain. The resolution is $(R,\phi,Z) = (256,1232,256)$, time step $\rm dt = \SI{1.62}{\, ns}$, plasma parameters are: $D_\perp = \SI{100}{\,m^2 s^{-1}}$, $\nu_0 = \SI{10}{\,m^2 s^{-1}}$, $\eta_0 = \SI{1.0E-3}{\,\Omega m}$, $\chi_\perp = \kappa_\perp/n = \SI{100}{\,m^2 s^{-1}}$ and $\chi_\parallel = \kappa_\parallel/n = \SI{1E3}{\,m^2 s^{-1}}$. Particle diffusivity, viscosity, resistivity, and perpendicular heat conductivity are larger than realistic values for simplicity of the study. Parallel heat diffusivity is of lower order than realistic values for the same reason. The summary of normalization of MHD system of the work is shown in the table \ref{tab:variables-denormalization} in \ref{Appx:denormalization-of-variables}.

\begin{figure}
    \centering
    \includegraphics[width=\linewidth]{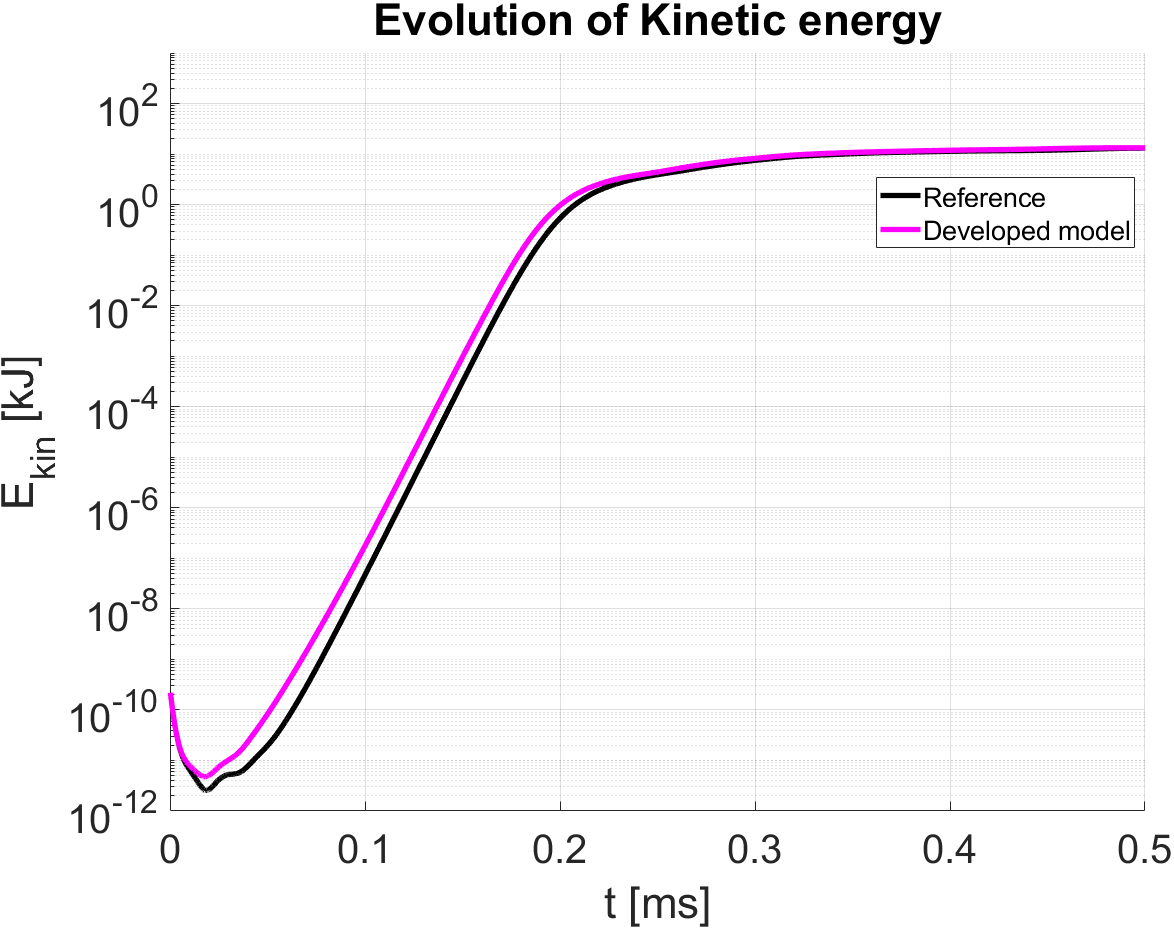}
    \caption{Kinetic energy evolution of IDB-SDC LHD $R_{\rm axV}  = \SI{3.85}{\,m}$ $\beta_{\rm 0}=4\%$ plasma for benchmark of the developed model (pink) against the reference model (black).}
    \label{fig:MIZUGUCHI-benchmark-energy}
\end{figure}
Figure \ref{fig:MIZUGUCHI-benchmark-energy} shows the time evolution of the kinetic energies of the plasma using the reference and developed MHD models. It has been verified that, using the same plasma parameters, the developed model can reproduce similar evolution of kinetic energy and energy growth rate as the former model. 
\begin{figure}
    \centering
    \includegraphics[width=\linewidth]{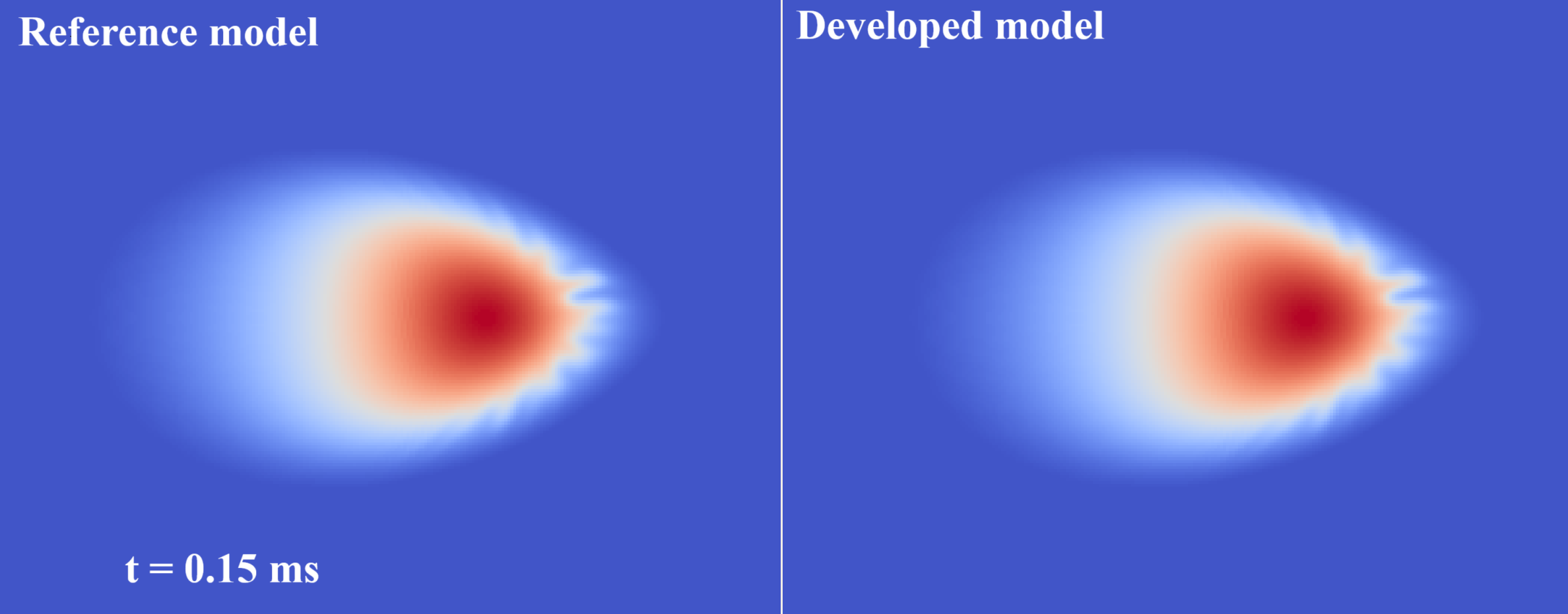}
    \caption{2D contour plot of pressure at $t = \SI{0.15}{\,ms}$ of IDB-SDC LHD $R_{\rm axV}  = \SI{3.85} {\,m}$ $\beta_{\rm 0}=4\%$ plasma for benchmark of the reference model (left) against the developed model (right).}
    \label{fig:MIZUGUCHI-benchmark-profile}
\end{figure}
Figure \ref{fig:MIZUGUCHI-benchmark-profile} shows the pressure profile for the developed model and the reference model at $t=\SI{0.15}{\,ms}$ (linear regime). The structures observed in both models are in agreement, with the perturbation localized in the low-field side of the plasma.

\section{Simulation results} \label{Sec:simulation-results}
The 3D MHD equilibrium is calculated using the HINT code, which considers the stochastic region of the magnetic field and resolves magnetic islands due to the non-linear 3D equilibrium response \cite{Suzuki_2006, Suzuki_2017}. The properties of the constructed equilibrium are discussed in Section \ref{Sec:3D-equilibrium}. 
The 3D non-linear MHD evolution is calculated by MIPS code based on the 3D equilibrium, and the analysis of the spontaneous CDC event, $i.e.$ the instability intrinsic to the configuration without any external source, is discussed in this section.

The resolution of the simulation is $(R,\phi,Z) = (384,1792,384)$. The time step is: $\rm dt = \SI{1.62}{\, ns} $. The values of parameters are: $D_\perp = \SI{1}{\,m^2 s^{-1}}$, $\nu_0 = \SI{100}{\,m^2 s^{-1}}$, $\eta_0 = \SI{1.0E-4}{\,\Omega m}$, $\chi_\perp = \kappa_\perp / n = \SI{1}{\,m^2 s^{-1}}$ and $\chi_\parallel = \kappa_\parallel / n = \SI{1E5}{\,m^2 s^{-1}}.$ Viscosity, resistivity and parallel heat conductivity are temperature dependent as: $\nu(T) \propto \nu_{\rm 0} (T/T_0)^{-3/2}$,  $\eta(T) \propto \eta_{\rm 0}(T/T_0)^{-3/2}$ and $\kappa_\parallel(T) \propto \kappa_{\parallel,\rm 0} (T/T_0)^{5/2}$, respectively, where $T_0$ is the equilibrium temperature at the magnetic axis. In the IDB particle diffusivity has been observed to be low \cite{Sakamoto_2008} but it has not been reproduced in this study. More realistic profiles of the IDB properties should be studied in future works.  The ratio $\kappa_\parallel / \kappa_\perp = \SI{1E5}{}$ is one order of magnitude smaller than what is expected ($\kappa_\parallel / \kappa_\perp = \SI{1E6}{}$) due to numerical limitations. The $\eta, \nu$ parameters used in this work are larger, and $\kappa_\parallel$ is smaller, than the realistic values because of numerical capability of the code.

\begin{figure}[t]
    \centering
    \includegraphics[width=\linewidth]{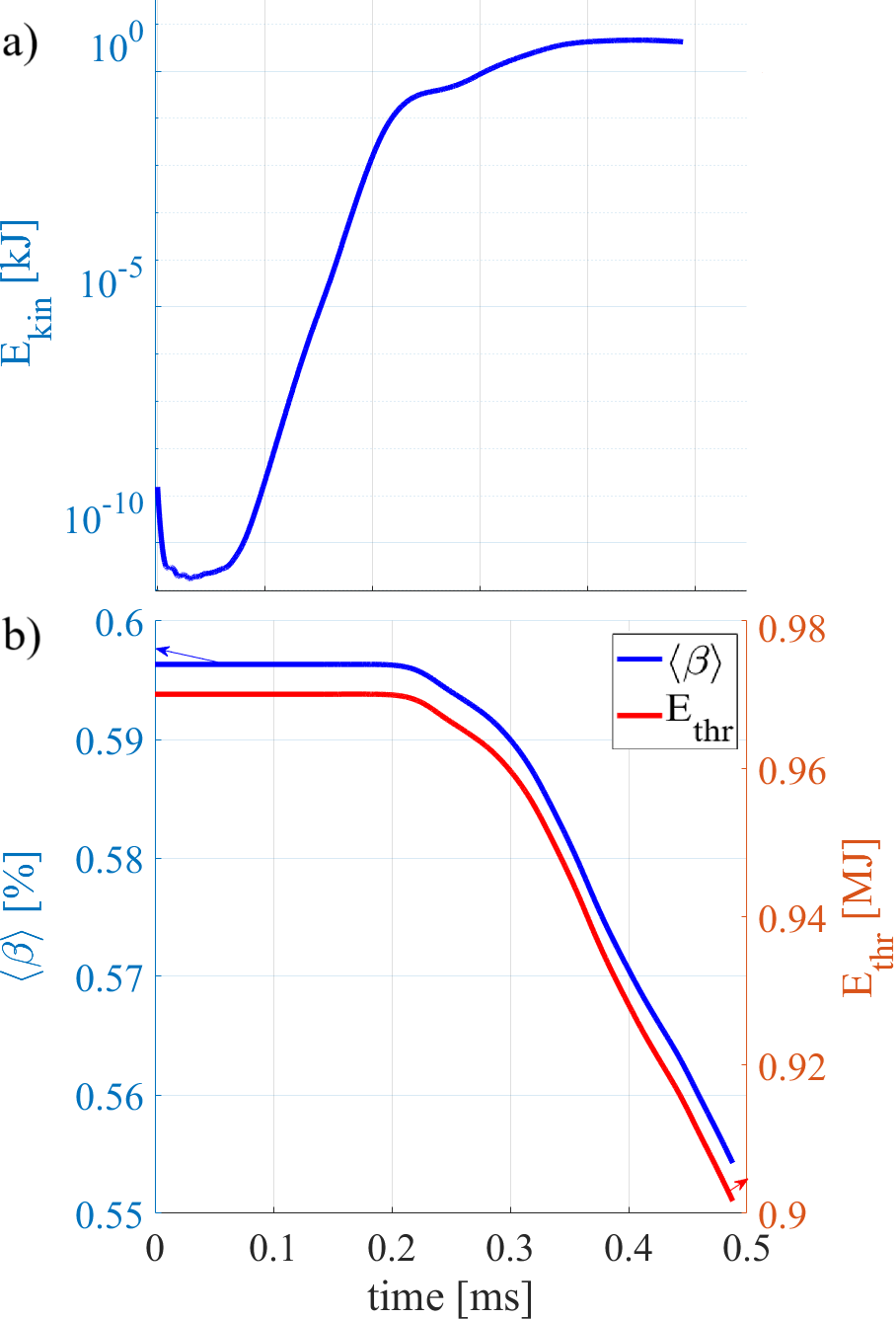}
    \caption{Evolution of a) the kinetic energy, and b) averaged beta $\langle\beta\rangle$ (blue - left axis) and thermal energy (red - right axis) of the CDC event in the IDB-SDC LHD $R_{\rm axV}  = \SI{3.85} {\,m}$ $\beta_0 = 4\%$ plasma}
    \label{fig:kinetic-energy-evolution-logaritmic}
\end{figure}
Figure \ref{fig:kinetic-energy-evolution-logaritmic}a) shows the time evolution of the total kinetic energy, $E_{kin} = \int (\rho v^2/2)\,\rm dV$ of the spontaneous CDC event case. The kinetic energy exponentially evolves during the linear stage, $t \sim 0.08 - \SI{0.22}{\, ms}$, then the energy saturates and the plasma reaches the relaxation stage,  $t > \SI{0.22}{\, ms}$. In Figure \ref{fig:kinetic-energy-evolution-logaritmic}b) the evolution of averaged $\beta$ value, $\langle\beta\rangle = 2\mu_0\langle p\rangle/\langle B^2\rangle, \quad$ and total thermal energy $E_{thr} = \int p/(\gamma-1)\, \rm dV$ is represented. The $\langle\beta\rangle(t=\SI{0.0}{\,s})$ value is smaller than $1\%$ since in MIPS code the volume integral is performed over the volume of binary mask, which is larger than the plasma region. Near the end of the linear regime $ t = \SI{0.22}{\, ms}$, the total confined energy and $\langle\beta\rangle$ decrease as the confinement is degraded. The subsequent $\langle\beta\rangle$ decrease is of $\sim 7\%$ in $t \sim \SI{0.3}{\,ms}$. Simulations have been stopped when plasma reaches the saturation of the ballooning modes, which is the scope of the work.
\begin{figure}[t]
    \centering
    \includegraphics[width=\linewidth]{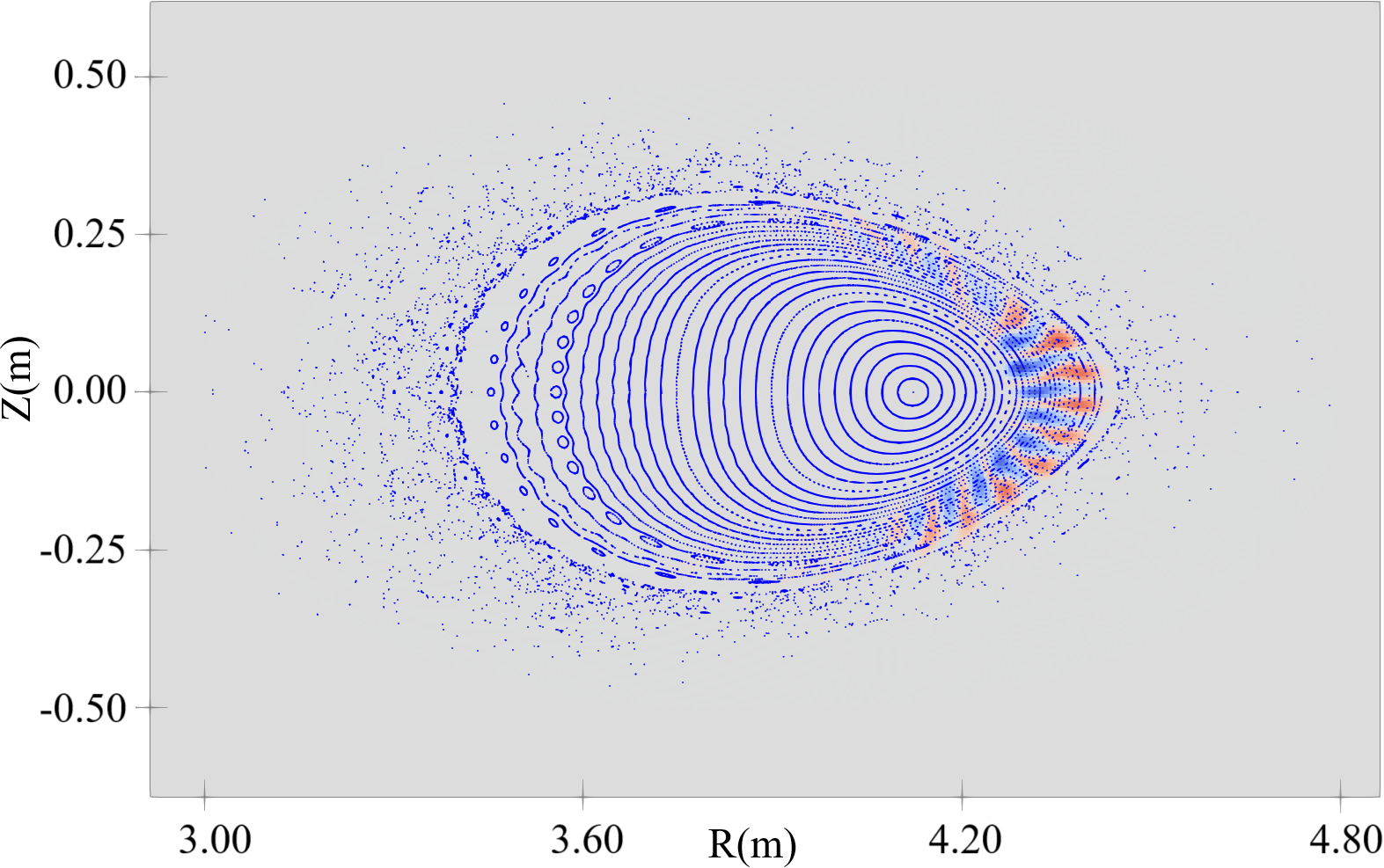}
    \caption{Perturbation of density in the horizontally elongated poloidal slice $\phi = (1/2)(2\pi/10)$ at $t = \SI{0.2}{\, ms}$. For reference, the Poincar\'e plot of the equilibrium magnetic field is represented in blue dots.}
    \label{fig:n-perturb_vs_Beq}
\end{figure}
Figure \ref{fig:n-perturb_vs_Beq} shows the mode structure of the perturbation of the density at $t = \SI{0.2}{\, ms}$. For reference the magnetic field topology of the equilibrium is added in the figure. The mode structure is localized in the low-field side of the plasma and has the structure of ballooning modes, which is in agreement with experimental observations \cite{Ohdachi_2017} and numerical simulations \cite{Mizuguchi_2008}. In the horizontally elongated poloidal slice, Thomson scattering diagnostics are utilized to measure electron temperature and density along the major radius \cite{Yamada_2010}. For this, the latter poloidal slice has been chosen for analysis in this study.

\begin{figure*}[t]
    \centering
    \includegraphics[width=\linewidth, keepaspectratio]{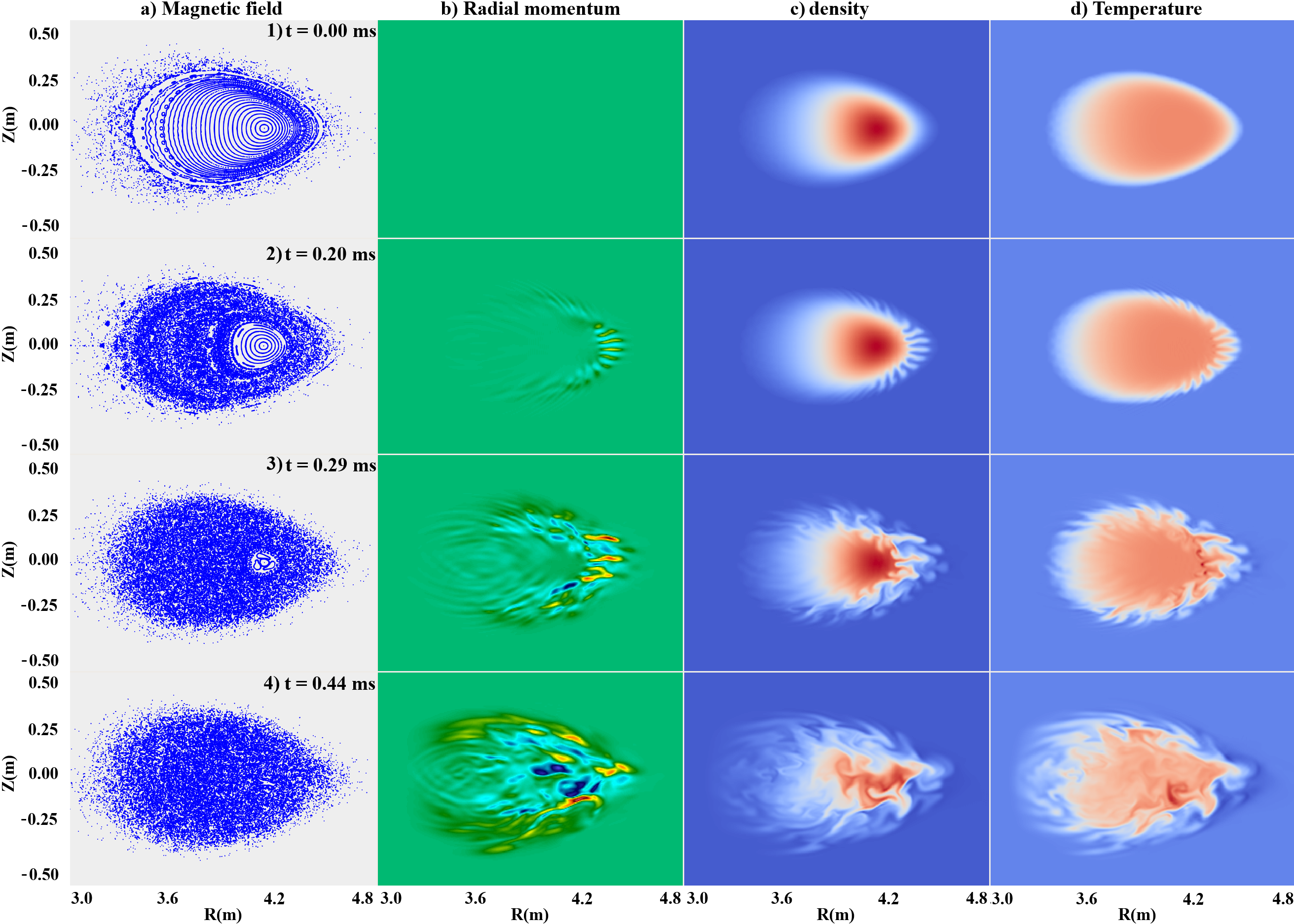}
    \caption{2D plots at $\phi=(1/2)(2\pi/10)$ of IDB-SDC LHD $R_{\rm axV}  = \SI{3.85} {\,m}$ $\beta_0 = 4\%$ plasma, for four different time slices, $t=\SI{0.00}{\, ms}$ (row 1), $t = \SI{0.20}{\,ms}$ (row 2), $t=\SI{0.29}{\,ms}$ (row 3) and $t=\SI{0.44}{\,ms}$ (row 4) of magnetic field (column (a)), radial momentum (column (b)), density (column (c)) and temperature (column (d)).}
    \label{fig:no-heat-source-2D}
\end{figure*}
Figure \ref{fig:no-heat-source-2D} shows the 2D contour plots at the horizontally elongated poloidal slice, $\phi=(1/2)(2\pi/10)$ of a) magnetic field, b) radial momentum, c) density, and d) temperature for four time slices. In the first row, $t=\SI{0.00}{\, ms}$, the equilibrium profiles of the plasma configuration are represented. The second row, $t=\SI{0.20}{\, ms}$, corresponds to the time slice near the end of the linear regime. In Fig. \ref{fig:no-heat-source-2D}(2-b) the radial momentum plot shows ballooning mode structures. The ballooning modes stochastize the magnetic field at the edge of the plasma while maintains the nested surfaces in the core, which is observed in Fig. \ref{fig:no-heat-source-2D}(2-a). The  density represented in Fig. \ref{fig:no-heat-source-2D}(2-c) and temperature in Fig. \ref{fig:no-heat-source-2D}(2-d) show perturbations localized in the low-field side due to the ballooning mode. The third row corresponds to $t = \SI{0.29}{\, ms}$, shortly after the plasma has reached the saturation regime. The ballooning mode structure observed in Fig. \ref{fig:no-heat-source-2D}(3-b) becomes wider than at the linear regime due to the saturation and relaxation of the modes, while the magnetic field, Fig. \ref{fig:no-heat-source-2D}(3-a), becomes more stochastic. The density, represented in Fig. \ref{fig:no-heat-source-2D}(3-c) and temperature profile, in Fig. \ref{fig:no-heat-source-2D}(3-d), are perturbed following the saturation of the ballooning modes. The perturbation is larger in the low-field side of the plasma, but spreads to the high-field side of the plasma and to the core. The fourth row, $t=\SI{0.44}{\, ms}$, corresponds to long after saturation regime is reached. The stochastization of the magnetic field, represented in Fig. \ref{fig:no-heat-source-2D}(4-a) remains after the saturation and increases with respect to \ref{fig:no-heat-source-2D}(3-a). The ballooning modes have saturated and spread further towards the high-field side of the plasma, represented in Fig. \ref{fig:no-heat-source-2D}(4-b). The density profile, Fig. \ref{fig:no-heat-source-2D}(4-c), has undergone the collapse and the plasma shape is significantly broken. The temperature profile, Fig. \ref{fig:no-heat-source-2D}(4-d), also shows significant perturbations, which is not in full agreement with the experimental results \cite{Ohdachi_2010, Ohdachi_2017, Suzuki_2021}.   
The evolution of the stochastic region is accompanied by an outward relaxation of the plasma and loss of confinement of plasma, consistent with the experimental observations of the CDC event. The modeling of the CDC crash in W7-X \cite{Suzuki_2021} is observed to completely stochastize the magnetic field  after the ballooning mode structure saturates, which occurs in this study, supporting the validity of the model.

\begin{figure}
    \centering
    \includegraphics[width=\linewidth]{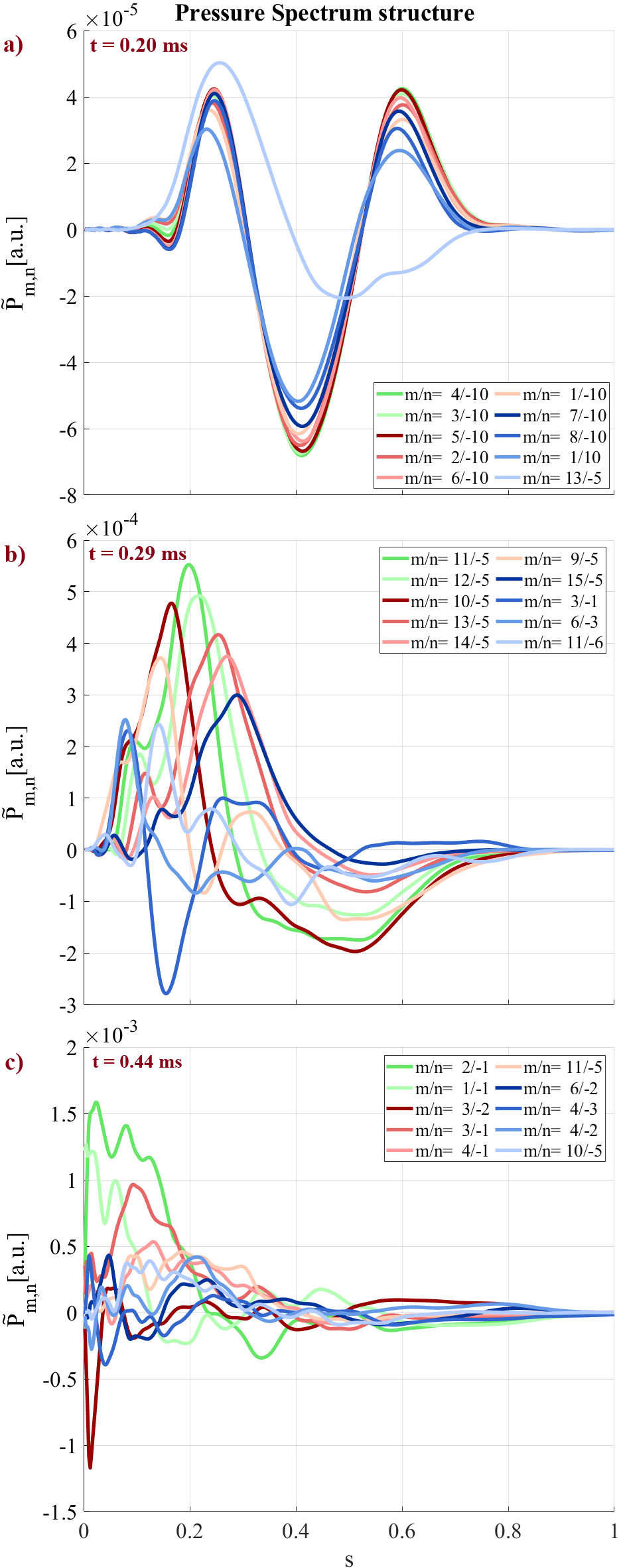}
    \caption{Radial profile of spectrum structure of pressure perturbation, $\tilde{P}_{\rm m,n}$, against normalized toroidal flux, $s$, of IDB-SDC LHD $R_{\rm axV}  = \SI{3.85} {\,m}$ $\beta_0 = 4\%$ plasma, sorted by strength of mode amplitude, for three different time slices, a) $t = \SI{0.20}{\,ms}$, b) $t=\SI{0.29}{\,ms}$  and c) $t=\SI{0.44}{\,ms}$.}
    \label{fig:mode-sprectum-prs}
\end{figure}
Figure \ref{fig:mode-sprectum-prs} shows the radial profiles of the mode structures of pressure perturbation, $\tilde{P}_{\rm m,n}$, from the pressure profile based on Figure \ref{fig:no-heat-source-2D}. The mode structure is represented as a function of the normalized toroidal flux, $s$. The toroidal and poloidal modes analyzed are $n= [-20, 20]$ and $m=[0,30]$, respectively. The dominant modes are represented in this figure sorted by magnitude. Three time slices are displayed in the figure, 1) $t = \SI{0.20}{\, ms}$, 2) $t = \SI{0.29}{\, ms}$, and 3) $t = \SI{0.44}{\, ms}$. During the linear phase (t = 0.20 ms) and the early nonlinear phase (t = 0.29 ms), toroidal modes appear predominantly at multiples of n = 5 and 10, reflecting the inherent periodicity of the LHD plasma. In the fully nonlinear stage (t = 0.44 ms), low-n modes become dominant, corresponding to the plasma reconfiguration induced by the CDC event. The mode structure shifts towards the core. This behavior is consistent with that observed in the W7-X core plasma collapse studied in \cite{Suzuki_2021}, which further supports the validity of the developed model, and points that the core collapse suffered in the LHD and W7-X follow similar MHD mode dynamics.

\begin{figure}[t]
    \centering
    \includegraphics[width=\linewidth]{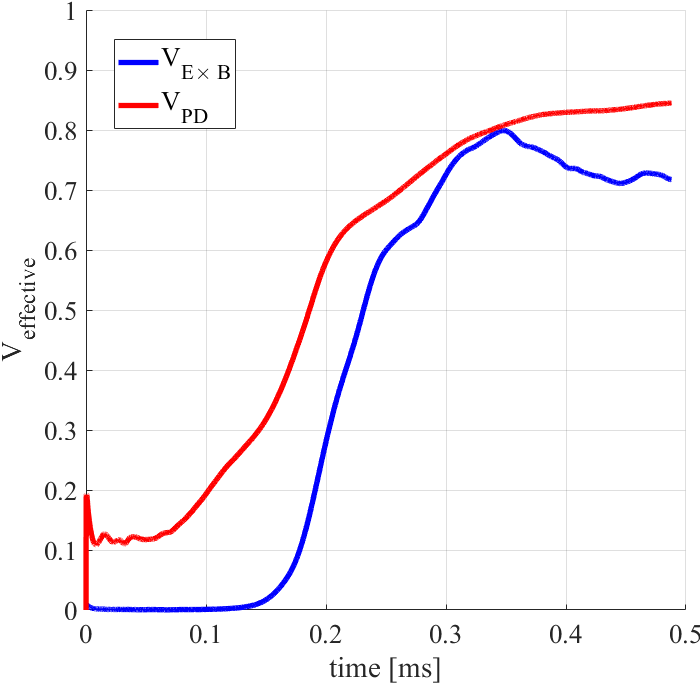}
    \caption{Evolution of the effective volumes $V_{\rm E\times B}$ and $V_{\rm PD}$ in the IDB-SDC LHD $R_{\rm axV}  = \SI{3.85} {\,m}$ $\beta_0 = 4\%$ plasma.}
    \label{fig:V-effective_evolution}
\end{figure}
Figure \ref{fig:V-effective_evolution} shows the evolution of the effective volume of $E\times B$ convection and effective volume of parallel diffusion. The effective volume of parallel diffusion, 
\begin{equation}
    V_{\rm PD} = \frac{1}{V}\int\mathcal{H}(\kappa_\parallel|\nabla_\parallel T|^2-\kappa_\perp |\nabla_\perp T|^2)\,\rm dV,
\end{equation}
where $\mathcal{H}$ is the Heaviside function, is computed in the simulation \cite{Paul_2022, Zhou_2024, Baillod_2023}. $V_{\rm PD}$ is dimensionless and ranges in $V_{\rm PD} \in [0,1]$. In a similar manner, the effective volume of $E\times B$ convection has been introduced comparing the heat convection, $v_{\rm E\times B}T$, due to $E\times B$ velocity, $\mathbf{v}_{\rm E\times B} = \mathbf{E}\times \mathbf{B}/B^2$; against heat conduction $|\chi_\parallel \nabla_\parallel T + \chi_\perp \nabla_\perp T|$. Then, the effective volume of $E\times B$ convection is defined as: 
\begin{equation}
    V_{\rm E\times B} = \frac{1}{V}\int \mathcal{H}(|V_{\rm E\times B}|T - |\chi_\parallel\nabla_\parallel T + \chi_\perp \nabla_\perp T|)\,\rm dV.
\end{equation}
The effective volume of parallel diffusion gives an idea of in what fraction of the plasma the parallel heat diffusion dominates over the perpendicular diffusion. In equilibrium, isotherms are aligned with the magnetic field; consequently, initially $V_{\rm PD} \simeq 0$. As the magnetic field topology becomes stochastic, $V_{\rm PD}$ is expected to grow.  Similarly, the effective volume of $E\times B$ convection gives an idea of in what fraction of the plasma the $E\times B$ heat convection is stronger than heat conduction. Initially, given the magnetostatic equilibrium the effective volume of $E\times B$ convection is expected to be low, $V_{\rm E\times B} \simeq 0$. The volume to compute the magnitudes is chosen to be that where $p > (1+\SI{1e-3}{})p_{edge}$. The effective volume of parallel diffusion starts slightly above $0$ since a small stochastic region of the magnetic field is included in the volume where the magnitudes are computed, which has a contribution to $V_{\rm PD}$. During the linear regime evolution of the plasma, $t \sim \SI{0.08}{}-\SI{0.22}{\, ms}$, $V_{\rm PD}$ and $V_{\rm E\times B}$ grow. The growth of the effective volume of parallel diffusion during the linear regime suggests that the magnetic field topology is stochastized by the ballooning modes and nested surfaces start to break, which is supported by Fig. \ref{fig:no-heat-source-2D}(2-a). Near the end of the linear regime, $t  = \SI{0.18}{\,ms}$, $V_{\rm PD} > 0.5$, which suggests that the parallel heat conduction dominates over the perpendicular heat conduction. At the saturation, $V_{\rm PD}$ grows continuously with a further stochastization of the magnetic field, in agreement with Figs.\ref{fig:no-heat-source-2D}(3-a,4-a). The evolution of the effective volume of $E\times B$ convection follows a similar behavior: during the linear regime, $t \sim \SI{0.08}{}-\SI{0.22}{\, ms}$,  $V_{\rm E\times B}$ grows, which indicates that the ballooning mode is exciting the $E\times B$ velocity. Close to $t  = \SI{0.22}{\,ms}$, at the end of the linear regime, $V_{\rm E\times B} > 0.5$, indicating that convection due to the $E\times B$ velocity is stronger than conduction. This dominance of $V_{\rm E\times B}$ at the end of the linear regime and afterwards suggests that the crash is mainly caused by convection rather than conduction. $V_{\rm PD}$ and $V_{\rm E\times B}$ both depend on the choice of parameters, $\chi_\parallel / \chi_\perp$, which affect the value of the effective volumes.

\begin{figure*}
    \centering
\includegraphics[width=\linewidth]{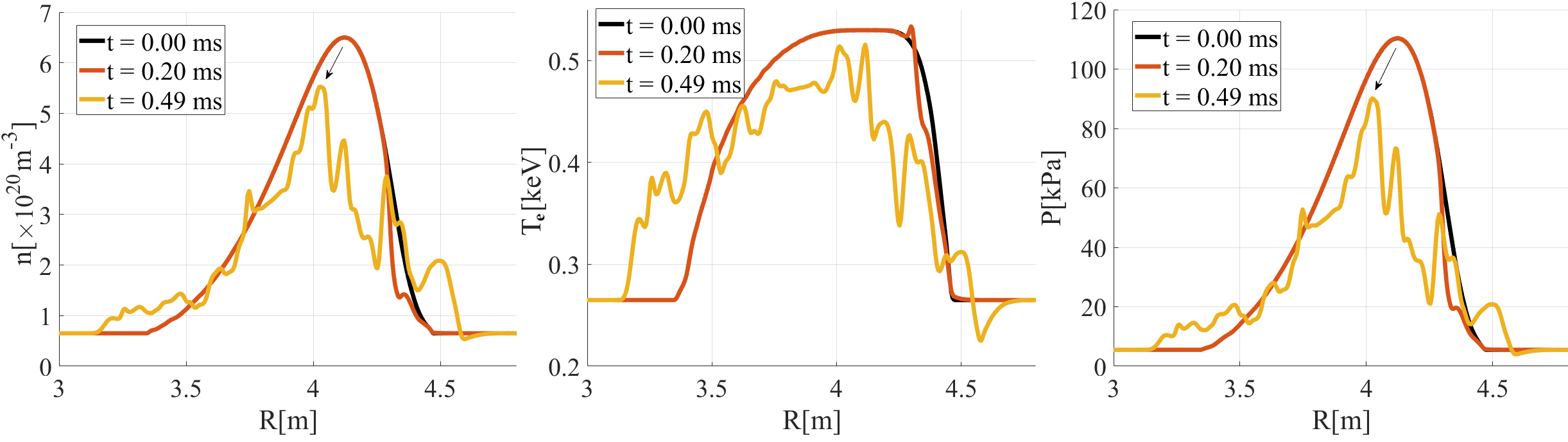}
    \caption{Radial profile versus major radius $(Z=\SI{0}{\,m}$, $\phi = (1/2)(2\pi/10))$ for density (left), electron temperature (center) and pressure (right) of the IDB-SDC LHD $R_{\rm axV}  = \SI{3.85} {\,m}$ $\beta_0 = 4\%$ plasma. Three time slices are represented, the equilibrium $t=\SI{0.00}{\,ms}$ (black), at the linear regime $t=\SI{0.20}{\,ms}$ (red) and at the relaxation stage $t=\SI{0.49}{\,ms}$ (orange). Arrows indicate the peak of density and pressure shift inwards at the relaxation stage.}
    \label{fig:no-heat-source-1D}
\end{figure*}
Figure \ref{fig:no-heat-source-1D} shows radial profiles versus major radius for density, temperature, and pressure for the three time slices: equilibrium ($t = \SI{0.0}{\,ms}$), near the end of the linear regime ($t = \SI{0.2}{\,ms}$) and at the relaxation stage ($t = \SI{0.49}{\,ms}$). The radial profiles are represented at the horizontally elongated poloidal slice, $\phi = (1/2)(2\pi/10)$ and $Z = \SI{0}{\,m}$ ($\simeq$ axis). 
At the linear regime, $t = \SI{0.2}{\,ms}$, the pressure and density at the edge of the plasma are perturbed, which corresponds to the mode structures induced by the ballooning modes observed in Fig. \ref{fig:no-heat-source-2D}(2-b,c,d). At the relaxation regime, the pressure and density collapse are observed, accompanied by the inward shift of the peak of pressure and density (indicated by arrows in the plot) due to the loss of $\langle\beta\rangle$ of the plasma, which is in agreement with experimental observations \cite{Ohdachi_2017}. The temperature in the core decreases and the profile is perturbed, which is not observed experimentally. The three profiles grow at the edges of the plasma and expand, which is in agreement with the plasma flushing observed experimentally in \cite{Yamada_2007,Ohdachi_2017}.
\begin{figure*}
    \centering
    \includegraphics[width=\linewidth]{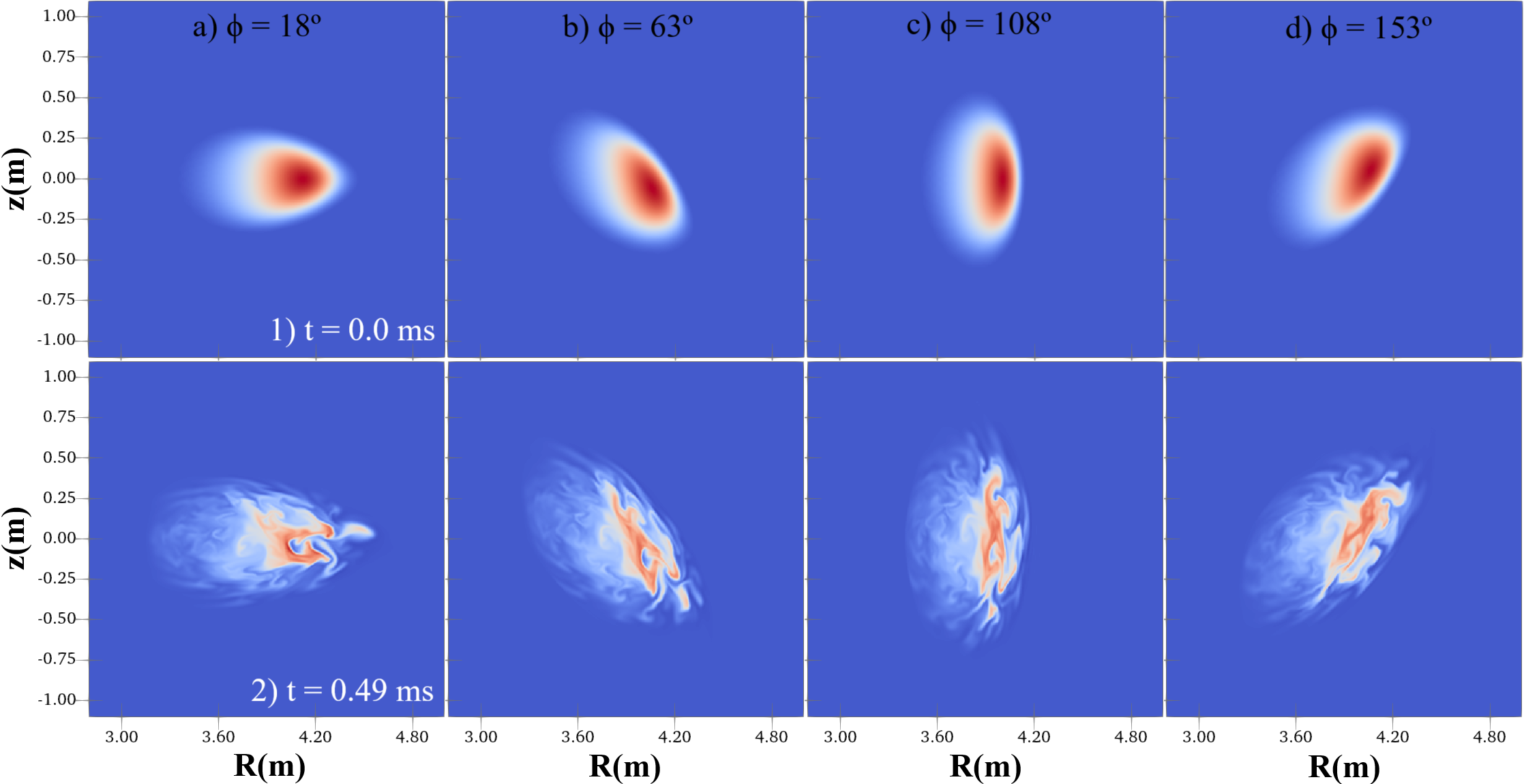}
    \caption{2D contour plots of density profile of the IDB-SDC LHD $R_{\rm axV}  = \SI{3.85} {\,m}$ $\beta_0 = 4\%$ plasma at the equilibrium (row 1) and at $t = \SI{0.49}{\,ms}$ (row 2) for four different poloidal slices of the reactor, $\phi = 18^\circ$ (column (a)), $\phi = 63^\circ$ (column (b)), $\phi = 108^\circ$ (column (c)) and $\phi = 153^\circ$ (column (d)).}
    \label{fig:2D-poloidal_slices_n_eq_vs_end}
\end{figure*}
Figure \ref{fig:2D-poloidal_slices_n_eq_vs_end} compares the equilibrium density 2D contour profile with the density at the relaxation stage $t=\SI{0.49}{\,ms}$ at four different poloidal slices throughout the torus. In the figure it is observed that the collapse of the density occurs at different toroidal angles.

The new model has been observed to reproduce the main characteristics of the CDC event i.e., the excitation of ballooning modes which lead to the collapse of the core density profile and pressure profile throughout the torus, together with the stochastization of the magnetic field.  In the simulation, temperature profile in the core is decreased and perturbed by the CDC event, which is not observed experimentally. Further improvement of the model and plasma set of parameters is required to be closer to the experimental observation for the future work.

\section{Conclusions and perspectives} \label{Sec:Conclusions-perspectives}
The developed $3$D non-linear MHD model extended in the MIPS code has been benchmarked and its validity has been confirmed against the previous version of MIPS code. 
The 3D non-linear MHD simulation shows preliminary results capturing the main characteristics of the CDC event observed in LHD plasma. Ballooning mode structures are observed in the outer region of the plasma before the CDC event. The ballooning modes have been observed to cause a core density collapse event, together with the stochastization of the magnetic field. At the linear regime, high-$n$ modes are dominant, but the non-linear coupling causes a shift to low-$n$ modes at the relaxation stage, which implies that the CDC event is not triggered by modes localized in the core purely, but is derived from high-$n$ ballooning modes localized in the edge, which then non-linearly couple and cascade into low-$n$ modes. The parallel heat conduction has been observed to increase as the ballooning modes develop, dominating over the perpendicular conduction ($V_{\rm PD} \gtrsim 0.5$) near the end of the linear regime, due to the increasing stochastization of the magnetic field. The $E\times B$ velocity convection is observed to grow with the ballooning modes, eventually overcoming the conduction ($V_{\rm E\times B} \gtrsim 0.5$) when plasma reaches the end of the linear regime, which indicates that $E\times B$ convection dominates rather than conduction. This observation insinuates that breaking of the plasma profile at the relaxation stage is caused by the $E\times B$ fluctuations induced by the ballooning modes rather than the diffusion due to the stochastization of the magnetic field. The CDC event causes the drop of the density profile and pressure profile in the core, and an increase of the profiles of density, temperature and pressure at the outer region of the plasma, replicating the flushing observed experimentally \cite{Ohdachi_2017}. Temperature profile decreases and loses the shape in the core during the simulation, which does not fully agree with experimental observations and may be caused by limitations of the model and plasma parameter set, which will be addressed in future work.  
Future studies are expected to extend the physics model and numerical model to allow performing with more realistic plasma parameters for comparison with experimental results.

\ack
One of the authors (Civit. A) has the support of the Joan Oró predoctoral grants program AGAUR-FI (2023 FI-3 00065) from the Secretary of Universities and Research from the Research and Universities Department from the Government of Catalonia and from European Social Fund Plus. 
This work received funding from the Spanish Ministry of Science (Grant No. PID2020-116822RBI00).
The authors gratefully acknowledge the computer resources through EUROfusion HPC project from Marconi-Fusion and Leonardo, the High Performance Computers at the CINECA headquarters in Bologna (Italy) and JFRS-1 provided under the EU-JA Broader Approach collaboration in the Computational Simulation Centre of International Fusion Energy Research Centre (IFERC-CSC). This work is performed on ‘Plasma Simulator’ (NEC SX-Aurora TSUBASA) of NIFS with the support and under the auspices of the NIFS Collaboration Research (NIFS22KISS005, NIFS22KIST012). Views and opinions expressed are however those of the author(s) only and do not necessarily reflect those of the European Union or the European Commission.

\section*{References}
\providecommand{\newblock}{}


\clearpage
\onecolumn
\appendix
\section{Table of normalization of MIPS variables} \label{Appx:denormalization-of-variables}

\begin{table}[hb]
\caption{\label{tab:variables-denormalization}Definition of the variables used in MIPS code, and the normalization from/to SI units.}
\renewcommand{\arraystretch}{1.7} 
\begin{tabular}{@{}lcccc}
\br
\textbf{Magnitude}            & \textbf{Variable}      & \textbf{Unit}                & \textbf{Dimensions}                 & \textbf{Normalization} \\
\br
Length               & $R, \vec{r}$  & $[m]$               & $[L]$                      & $\hat{r}$\\
Magnetic field       & $\vec{B}$     & $[T]$               & $[M T^{-2} I^{-1}]$        & $B_0\hat{B}$\\
Density              & $n$           & $[m^{-3}]$          & $[L^{-3}]$                 & $n_0\hat{n}$\\
Mass density         & $\rho$        & $[kg/m^3]$          & $[M L^{-3}]$               & $\rho_0\hat{\rho}$\\
Pressure             & $p$           & $[N/m^2]$           & $[ML^{-1}T^{-2}]$          & $\frac{B_0^2}{\mu_0}\hat{p}$\\
Energy (Temperature) & $T$           & $[J]$ or $[eV]$     & $[ML^{2}T^{-2}]$           & $\frac{B_0^2}{\mu_0n_0}\hat{T}$ or $\frac{B_0^2}{Z_e\mu_0n_0}\hat{T}$ \\
Velocity             & $\vec{v}$     & $[m/s]$             & $[LT^{-1}]$                & $v_A\hat{v} = \frac{B_0}{\sqrt{\mu_0 \rho_0}}\hat{v}$\\
Time                 & $t$           & $[s]$               &  $[T]$                     & $\tau_A\hat{t}=R_0/v_A\hat{t}$\\
Current density      & $\vec{J}$     & $[A/m^2]$           & $[IL^{-2}]$                & $\frac{B_0}{\mu_0R_0}\hat{J}$\\
Resistivity          & $\eta$        & $[\Omega \, m]$        & $[ML^3T^{-3}I^{-2}]$       & $\frac{\mu_0R_0^2}{\tau_A}\hat{\eta}$\\
Kinematic viscosity  & $\nu$         & $[m^2/s]$           & $[L^{2}T^{-1}]$            & $\frac{R_0^2}{\tau_A}\hat{\nu}$\\
Particle diffusivity & $D_\perp$     & $[m^2/s]$           & $[L^{2}T^{-1}]$            & $\frac{R_0^2}{\tau_A}\hat{D}$\\
Heat diffusivity     & $\chi$        & $[m^2/s]$           & $[L^{2}T^{-1}]$            & $\frac{R_0^2}{\tau_A}\hat{\chi}$\\
Particle source      & $S_{\rm N}$   & $[m^{-3}s^{-1}]$    &$[L^{-3}T^{-1}]$            & $\frac{n_0}{\tau_A}\hat{S}_{\rm N}$\\
Momentum source      & $\vec{S}_{\rho\vec{v}}$   & $[kg\, m^{-2}s^{-2}]$    &$[ML^{-2}T^{-2}]$            & $\frac{\rho_0 v_A} {\tau_A}\hat{S}_{\rho\vec{v}}$\\
Energy source      & $S_{\rm T}$     & $[W/m^3]$ or $[eV/m^3s]$ & $[ML^{-1}T^{-3}]$     & $\frac{B_0^2}{\mu_0\tau_A}\hat{S}_{\rm T}$ or $\frac{B_0^2}{Z_e\mu_0\tau_A}\hat{S}_{\rm T}$ \\
\br
\end{tabular}
\end{table}

In this work, the reference values used are: $B_0=\SI{2.77}{\,T}$, $R_0=\SI{3.85}{\,m}$, $n_0=\SI{6.5e+20}{\,m^{-3}}$, $\rho_0= n_0 \times m_p~\mathrm{[kg \cdot m^{-3}]}$, $m_p=\SI{1.67262192E-27}{\,kg}$, $\mu_0= 4\pi \times\SI{e-7}{\,Hm^{-1}}$, $Z_{\rm e}=\SI{1.60217663E-19}{\,C}$. The normalized values used in the numerical simulation are: Particle diffusivity $D_\perp = 10^{-7}$, viscosity $\nu_{0} = 10^{-5}$, resistivity $\eta_{0} = 10^{-5}$, perpendicular and parallel heat conductivity $\kappa_\perp = 10^{-7}$, $\kappa_\parallel = 10^{-2}$, respectively, with heat diffusivity $\chi \propto \kappa / n$.

\end{document}